\documentclass[aps,prl,twocolumn,showpacs]{revtex4}
\usepackage{epsfig}
\begin{document}
\title{Stochastic volatility and leverage effect}
\author{Josep Perell\'o and Jaume Masoliver\cite{email}}
\affiliation{Departament de F\'{\i}sica Fonamental, Universitat de Barcelona,\\ Diagonal, 647, E-08028 Barcelona, Spain}
\date{\today}

\begin{abstract}
We prove that a wide class of correlated stochastic volatility models exactly measure an  empirical fact in which past returns are anticorrelated with future volatilities: the so-called ``leverage effect''. This quantitative measure allows us to fully estimate all parameters involved and it will entail a deeper study on correlated stochastic volatility models with practical applications on option pricing and risk management.
\end{abstract}
\pacs{89.65.Gh, 02.50.Ey, 05.40.Jc, 05.45.Tp}
\maketitle

The multiplicative diffusion process known as the geometric Brownian motion (GBM) has been widely accepted as one of the most universal models for speculative markets. The model, started out by Bachelier in 1900 as an ordinary random walk and redefined in its final version by Osborne in 1959~\cite{cootner}, presupposes a constant ``volatility'' $\sigma$ which is equivalent to a constant diffusion coefficient $D=\sigma^2$. However, and especially after the 1987 crash, there seems to be ample empirical evidence, given by the so-called ``stylized facts'', that the assumption of constant volatility does not properly account for important features of markets. Some relevant examples of stylized facts are the ``leverage'' and ``smile'' effects and the skewness and ``fat tails'' in probability distributions~\cite{cont}. 

One of the main ideas that has come out to explain those features is that the market dynamics has intrinsically changed in the sense that volatility is no longer constant. It is not a function of time either (as might be inferred by the evidence of nonstationarity in financial time series~\cite{lo}) but a {\it random variable}. In its most general form one therefore assumes that the volatility $\sigma$ is a function of a random process $Y(t)$, {\it i.e.}, $\sigma(t)=\phi(Y(t))$. We could make an analogy from physics saying that speculative prices $S(t)$ evolve in a ``random medium'' determined by a random diffusion coefficient. Most of the stochastic volatility (SV) models presented up to date suppose that $Y(t)$ is itself a diffusion process that may or may not be correlated with price and different models differ from each other basically in the form of the function $\phi$ \cite{ghysels,fouquebook}. 

As mentioned above, the hypothesis of stochastic volatility (SV) has been suggested to explain significant peculiarities observed in real markets. Thus, the smile effect, related to the implicit volatility in option prices, has been thoroughly studied both qualitatively and quantitatively~\cite{smile,heston}. Other features such as leverage and long tails are less studied and some works on SV only address to them from a qualitative point of view~\cite{fouque} while others measure them by giving numerical coefficients, based on ARCH-GARCH models, for kurtosis and skewness~\cite{englepat}. However, the time evolution and structure of the leverage and tails have never been investigated. Our main objective here is to prove that the leverage effect can be quantitatively explained by a wide class of correlated SV models. This will allow us to overturn the main objection against SV models: the impossibility of fitting all parameters appearing in these models 
\cite{fouquebook,fouque} which, in turn, opens the door to simulations of real markets with far reaching practical  consequences on option pricing and risk management.

We recall that the leverage effect has its origin in the observation that volatility seems to be negatively correlated with stock returns which, in continuous time finance and in terms of speculative prices $S(t)$, are defined by $R(t)=\ln[S(t)/S(0)]$. The first explanation to this empirical fact was given by Black \cite{black} and Christie \cite{christie} in the sense that negative returns increase financial leverage which extend the risk of the company and therefore its volatility. Hence the name of ``leverage effect''. Nevertheless, the cause of this effect is still unclear since another explanation is just the contrary, that is, an increase of volatility makes the stock riskier which produces a fall of demand and the price drops \cite{ghysels}.

In a very recent paper, Bouchaud {\it et al.} \cite{bouchaud} have performed a complete empirical study of the leverage effect for both individual stocks and indices using daily data. The volatility-return (negative) correlation is clearly shown to have a definite direction in time -- a very confusing fact in the literature -- since correlations are shown to be between future volatilities and past returns. Bouchaud {\it et al.} conclusively prove  from data that the negative correlation decays exponentially in time, faster for indices than for individual stocks \cite{bouchaud}. 

In this letter, we present a theoretical study on these correlations and show that a wide class of SV models completely explain the leverage effect, both qualitatively and quantitatively, in complete agreement with experimental observations. The starting point is the GBM model:
\begin{equation}
dR=\mu dt+\sigma(t)dW_1(t),
\label{gbm}
\end{equation}
where $\mu$ is the drift and $\sigma(t)=\sigma(Y(t))$ is a random volatility and $Y(t)$ is a diffusion process:
\begin{equation}
dY=f(Y)dt+g(Y)dW_2(t).
\label{Y}
\end{equation}
In these equations $W_i(t)$ $(i=1,2)$ are Wiener processes, {\it i.e.,} $dW_i(t)=\xi_i(t)dt$ where $\xi_i(t)$ are zero-mean Gaussian white noises with 
cross-correlation given by
\begin{equation}
\langle\xi_1(t)\xi_2(t')\rangle=\rho\delta(t-t'),
\label{correl}
\end{equation}
$(-1\leq\rho\leq 1)$. As is common in finance, Eqs. (\ref{gbm})-(\ref{Y}) are interpreted in the sense of Ito and for the rest of the paper we will follow Ito convention \cite{perello1}. 

Bouchaud {\it et al.}~\cite{bouchaud}, quantify the leverage effect by means of a leverage correlation function defined by
\begin{equation}
{\cal L}(\tau)\equiv
\frac{1}{Z}\langle[dX(t+\tau)]^2dX(t)\rangle
\label{leverage}
\end{equation}
where 
\begin{equation}
X(t)\equiv R(t)-\mu t
\label{zeromean}
\end{equation}
is the zero-mean return and $Z=[\langle dX(t)^2\rangle]^2$ is a convenient normalization coefficient. Bouchaud {\it et al.} have analyzed a large amount of daily relative changes for either market indices and stock share prices and find that~\cite{bouchaud}
\begin{equation}
{\cal L}(\tau)=\cases{-A e^{-b\tau}, &if $\tau>0$;\cr
0, &if $\tau<0$;}
\label{bouchaudlev}
\end{equation}
$(A,b>0)$. Hence, there is a negative correlation with an exponential time decay between future volatility and past returns changes but no correlation is found between past volatility and future price changes. In this way, they provide a sort of causality to the leverage effect which, to our knowledge, has never been previously mentioned in the literature~\cite{ghysels,fouquebook}.

Let us sketch how correlated SV models are able to exactly reproduce this result. We first combine Eqs. (\ref{gbm}) and (\ref{zeromean}) to produce ${\cal L}(\tau)=\langle\sigma(t)dW_1(t)\sigma(t+\tau)^2dW_1(t+\tau)^2\rangle/Z$. Note that when $\tau<0$ Ito's rules tell us that $dW_1(t)$ is uncorrelated with the rest of terms then, recalling that $\langle dW_1(t)\rangle=0$, we have ${\cal L}(\tau)=0$ if $\tau<0$. On the other hand, when $\tau>0$, $dW_1(t+\tau)$ is uncorrelated with the rest and, since $\langle dW_1(t+\tau)\rangle^2=dt$, we conclude that
\begin{equation}
{\cal L}(\tau)=\theta(\tau)\langle\sigma(t)dW_1(t)\sigma(t+\tau)^2\rangle dt/Z,
\label{leverage1}
\end{equation}
where $\theta(\tau)$ is the Heaviside step function and 
$Z=[\langle\sigma^2(t)\rangle]^2dt^2$. Note that we have proved the existence of correlations between future volatilities and past returns but not vice-versa. Note also that this have been proved independently of the underlying volatility process $Y(t)$ 
(which needs not to be a diffusion process) and of the specific form of $\sigma$ in terms of $Y$. 

Suppose now that $Y(t)$ is a diffusion process given by Eq. (\ref{Y}). As is well known any pair of correlated Wiener process, such as $W_1(t)$ and $W_2(t)$, satisfy the identity $dW_1(t)=\rho dW_2(t)+\sqrt{1-\rho^2}dW(t)$, where $W(t)$ is a Wiener process independent of $W_2(t)$ (therefore, $W(t)$ is independent of $\sigma$). Substituting this identity into Eq. (\ref{leverage1}) we get ${\cal L}(\tau)$ in terms of the average 
$\langle\sigma(t)\sigma(t+\tau)^2\xi_2(t)\rangle$. This average can be calculated by means of Novikov's theorem \cite{novikov} with the result \cite{pm}
\begin{equation}
{\cal L}(\tau)=\frac{2\rho\theta(\tau)}{[\langle\sigma^2(t)\rangle]^2}
\left\langle\sigma(t)\sigma(t+\tau)\sigma'(t+\tau)
\frac{\delta Y(t+\tau)}{\delta\xi_2(t)}\right\rangle,
\label{leverage2}
\end{equation}
where $\sigma'=\partial\sigma(Y)/\partial Y$ and $\delta Y(t+\tau)/\delta\xi_2(t)$ is the functional derivative of $Y(t+\tau,[\xi_2])$ with respect to $\xi_2(t)$ \cite{novikov}.

\begin{figure}[t,b,h]
\centerline{\epsfig{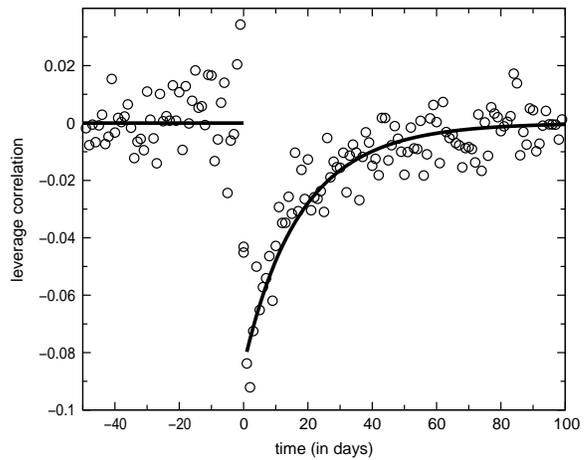}}
\vspace*{13pt}
\caption{The leverage effect in the Dow-Jones daily index. We plot the leverage function ${\cal L}(\tau)$ for the Dow-Jones index from 1900 until 2000. We see that there exists a non-negligible correlation when $\tau>0$ and negligible when $\tau < 0$. Observe that correlation strongly fluctuates when $-3<\tau<2$. We also plot a fit with the OU SV leverage function~(\ref{L1}) that helps us to estimate $\alpha$ and $\rho$.}
\label{djleverage}
\end{figure}

There is a wide consensus on volatility being ``mean reverting''. This means that there exists a normal level of volatility \cite{footnote} to which volatility will eventually return \cite{englepat}. For a general SV model such as 
(\ref{gbm})-(\ref{Y}), the presence of mean-reversion implies restrictions on the form of the drift coefficient $f(Y)$. In order to include this experimental fact in the model, the simplest choice  is to assume that $f(Y)$ is a linear function. That is, 
\begin{equation}
\dot{Y}=-\alpha(Y-m)+g(Y)\xi_2(t),
\label{Y2}
\end{equation}
where $\alpha<0$. The formal solution to this equation in the stationary state reads
$$
Y(t)=m+\int_{-\infty}^{t}e^{-\alpha(t-t')}g(Y(t'))\xi_2(t')dt',
$$
from which we get \cite{pm}
$$
\frac{\delta Y(t+\tau)}{\delta\xi_2(t)}=\theta(\tau)e^{-\alpha\tau}g(Y(t))
\exp\left[\int_t^{t+\tau}g'(Y(s))\xi_2(s)ds\right].
$$
Substituting this expression into Eq. (\ref{leverage2}) yields 
\begin{equation}
{\cal L}(\tau)=\rho\theta(\tau)B(\tau)e^{-\alpha\tau},
\label{leverage3}
\end{equation}
where
\begin{equation}
B(\tau)=\frac{2\left\langle\sigma(t)\sigma(t+\tau)\sigma'(t+\tau)G(t,t+\tau)\right\rangle}
{[\langle\sigma^2(t)\rangle]^2},
\label{B}
\end{equation}
and
\begin{equation}
G(t,t+\tau)=g(Y(t))\exp\left[\int_t^{t+\tau}g'(Y(s))\xi_2(s)ds\right].
\label{G}
\end{equation}
In consequence, any SV model of the form given by Eqs. (\ref{gbm}) and (\ref{Y2}) and whose  function $B(\tau)$ does not increase faster than $e^{\alpha\tau}$ as $\tau\rightarrow\infty$, satisfies an exponentially decaying leverage as expressed by 
Eq. (\ref{bouchaudlev}). Moreover, if $\sigma(Y)$ is an increasing function of $Y$ with definite sign and $g(Y)$ is positive definite (or $\sigma$ is decreasing and $g$ is negative) we see  from Eq. (\ref{leverage3}) that the correlation coefficient $\rho$ must be negative and  driving noises $W_1(t)$ and $W_2(t)$ are  anticorrelated. 
Eqs. (\ref{leverage3})-(\ref{G}) constitute the main result of the paper. 

The exact form of ${\cal L}(\tau)$ will depend on the expression of $B(\tau)$ which in turn  will depend on the SV model chosen. Within diffusion theory, as is the case of 
Eq. (\ref{Y2}), there are basically three different SV models \cite{fouquebook}: 1) The Ornstein-Uhlenbeck (OU) model where $\sigma=Y$ and $g(Y)=k$ (a positive constant) \cite{stein}, 2) the exponential Ornstein-Uhlenbeck (expOU) model where $\sigma=e^Y$ and $g(Y)=k$ \cite{fouque} and 3) the Cox-Ingersoll-Ross (CIR) model where $\sigma=\sqrt{Y}$ and $g(Y)=k\sqrt{Y}$ \cite{heston}. For all these models the leverage function has the form given by Eq. (\ref{leverage3}). In the OU model and in the CIR model (the latter with zero mean-reversion, {\it i.e.}, $m=k^2/4\alpha$) the leverage function ${\cal L}(\tau)$ is respectively given by \cite{pm}
\begin{equation}
{\cal L}_{1}(\tau)=2k\rho\theta(\tau)\left[\frac{m^2+(k^2/2\alpha)e^{-\alpha\tau}}
{(m^2+k^2/2\alpha)^2}\right]e^{-\alpha\tau}, 
\label{L1}
\end{equation}
\begin{equation}
{\cal L}_{3}(\tau)=4\frac{\rho\alpha}{k}\theta(\tau)e^{-\alpha\tau},
\label{L3}
\end{equation}
while for the expOU model we have
\begin{equation}
{\cal L}_{2}(\tau)=
2\rho k\theta(\tau)\exp\left[-m+k^2(e^{-\alpha\tau}-3/4)/\alpha\right]e^{-\alpha\tau}.
\label{L2}
\end{equation}
In Fig.~\ref{djleverage} we show the leverage effect for the Dow-Jones daily index (1900-2000) and plot the leverage function for the OU model. 

Any market model, besides being able to reproduce the market dynamics, must provide a systematic way of evaluating its parameters. Almost all current SV models have four parameters to estimate: $\rho$, $k$, $m$, and $\alpha$. To our knowledge, all works on SV models presented up till now are able to evaluate only two of them. Thus, for instance, Fouque {\it et al.} \cite {fouque} estimate $k$ and $m$ from the empirical second and fourth moment of daily data but they can't give a clear estimation of $\alpha$ and $\rho$. This constitutes the main criticism to SV models, say, their inability to estimate all parameters involved. This situation changes completely when one measures leverage. Indeed, $k$ and $m$ are obtained, as usual, from the empirical second and fourth moment. Next, by adjusting $e^{-\alpha\tau}$ to leverage  empirical data we estimate $\alpha$. Moreover, comparing the theoretical and empirical leverage at $\tau=0^+$, ${\cal L}(0^+)$, we finally obtain $\rho$. Following this procedure, for the Dow-Jones daily index and using the OU model, we have estimated $k=1.4\times 10^{-3} \mbox{ days}^{-1}$, $m=18.9 \%\mbox{ year}^{-1/2}$, $\alpha= 0.05 \mbox{ day}^{-1}$, and $\rho=-0.58$. 

\begin{figure}[t,b,h]
\vspace*{13pt}
\centerline{\epsfig{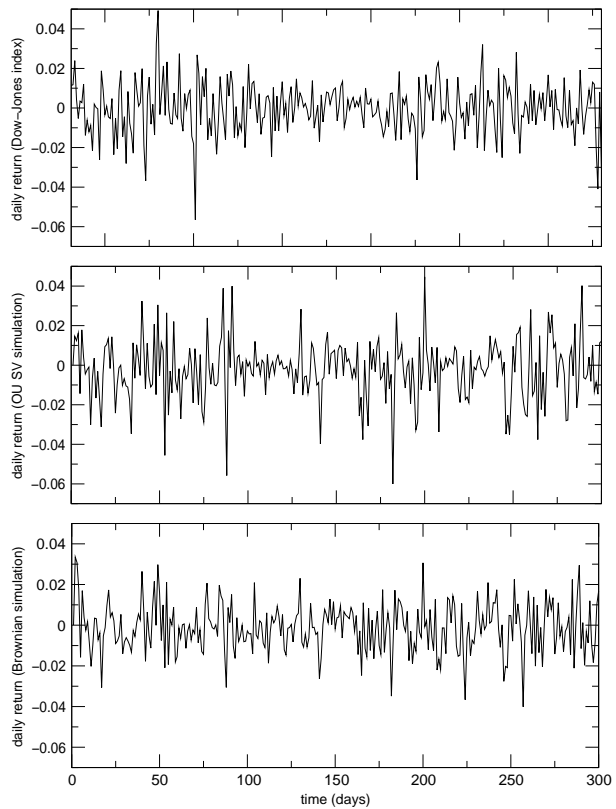}}
\vspace*{13pt}
\caption{Dow-Jones historical time-series (1900-2000) and path simulations. We show a Dow-Jones daily returns sample path (top), its simulation by means of the OU SV process 
(middle), and the geometric Brownian process simulation (bottom). All parameters of the simulations are estimated from the whole Dow-Jones historical time-series from 1900 to 2000. The dynamics is traced over 300 days (approximately one trading year) which, for the empirical path, nearly corresponds to 1999.}
\label{comp}
\end{figure}

As an illustration we have simulated, using Eqs. (\ref{gbm}) and (\ref{Y2}), the OU resulting process with the parameters estimated above. We follow the random dynamics of the daily changes of the zero-mean return $X(t)$ and compare it with the empirical Dow-Jones time series during one trading year. We have also simulated the geometric Brownian motion assuming a constant volatility $\sigma$ whose value is directly estimated from Dow-Jones one-century data. We present these results in Fig.~\ref{comp} and observe that GBM cannot describe either the largest or the smallest fluctuations of daily returns. We nonetheless see in the figure that the SV model chosen describes periods of high volatility together with periods of very low volatility, resulting in a more similar trajectory to the Dow-Jones index than that of the GBM. This is quite remarkable, because we have simulated last year trajectory using all past data (one century) of the Dow-Jones index thus showing the stability of parameters. 

These results may encourage to deepen in the study of statistical properties of SV models showing leverage. Several models have been presented in the literature without being able to discern which is the more realistic one. Now, thanks to leverage correlation, it is possible to estimate all parameters involved in SV models. This will allow us to confront in detail different models with the empirical statistical properties of markets. Finally, a better knowledge of SV models has non trivial consequences on option pricing (since classical Black-Scholes method is still suitable in SV models \cite{smile,heston}) and, more generally, on risk management.

\acknowledgements
This work has been supported in part by Direcci\'on General de
Investigaci\'on under contract No. BFM2000-0795, by Generalitat de Catalunya under contract
No. 2000 SGR-00023. We thank Marian Bogunya for useful suggestions and careful reading of the manuscript.


\begin{references}
\bibitem{footnote} The normal level of volatility is usually defined by $\lim_{t\rightarrow\infty}\langle\sigma^2(t)\rangle$ (see \cite{englepat}).
\end{references}
\end{document}